# GEOMECHANICS IN UNCONVENTIONAL RESOURCE DEVELOPMENT

Binh T. Bui[1], PetroVietnam University, Ba Ria, Ba Ria-Vung Tau, Vietnam

**Abstract**

To economically produce from very low permeability shale formations, hydraulic fracturing stimulation is typically used to improve their conductivity. This process deforms and breaks the rock, hence requires the geomechanics data and calculation. The development of unconventional reservoirs requires large geomechanical data, and geomechanics has involved in all calculations of the unconventional reservoir projects. Geomechanics has numerous contributions to the development of unconventional reservoirs from reservoir characterization and well construction to hydraulic fracturing and reservoir modeling as well as environmental aspect. This paper reviews and highlights some important aspects of geomechanics on the successful development of unconventional reservoirs as well as outlines the recent development in unconventional reservoir geomechanics. The main objective is to emphasize the importance of geomechanical data and geomechanics and how they are being used in in all aspects of unconventional reservoir projects.

## Contribution of unconventional reservoir in hydrocarbon production

Shale has long been known as abundant caprock that has a significant amount of hydrocarbon. However, due to its nanopore size, the flow of hydrocarbons toward the wellbore is uneconomical without some advanced technologies. The rapid development of the unconventional resources is supported mainly by the combination of horizontal drilling and hydraulic fracturing. The first success of the shale hydraulic fracturing in shale by Mitchell Energy 1998 has started the unconventional resource revolution in the U.S. making the country more energy independence and gradually becoming an importer of oil and gas. While the Annual Energy Outlook 2014 predicted that the total projected U.S. crude oil production can reach 9.6 MMbbl/d in 2019, the daily production at the end of 2018 already exceeded 11.5 MMbbl/d (EIA 2018). The production from tight unconventional formations has grown rapidly from 2.5 MMbbl/d in 2012 (EIA 2014) to about 4.3 MMbbl/d in 2018 (EIA 2018). It is anticipated that the tight oil production will continue to reach 6 MMbbl/d in 2029 (EIA 2018). Shale gas reserves and production also grow rapidly and is becoming the dominant source of natural gas in the U.S. In 2000, shale gas only contributed only more than 1% of the U.S. dry gas production, today shale gas contributes more than 70% of U.S. dry gas production. The shale gas share of total U.S. natural gas production is projected to increase from 40% in 2012 to 53% in 2040 (EIA 2014). Geomechanics is one of the important factors contributing to the success of unconventional reservoir development since it involves in all aspects of operations from drilling to production as well as many other aspects such as environmental protection and sustainability development.

## Unconventional reservoir geomechanics

Unconventional reservoir is used to refer to the reservoir that has very low permeability that requires stimulation techniques to produce economically. Shale, is the most important part of unconventional resource, is sedimentary mainly composed muds, silts and clay that has particle size is in the range of 0.0625 - 0.004 mm with relatively high organic content. The main composition of shale are carbonates, quartz, feldspar, clay and organic content (**Figure 1**). Shale is often considered as the source rock for the formation of hydrocarbon during the maturation process. Because of its very fined particle size, shale matrix permeability is very low, typically in the range of nano-Darcy while the permeability of the natural fracture is in the range of micro-Darcy. From the transport point of view, shale has very low matrix and natrual fracture permeability and porosity, and the dependece of these transport properties on stress is significant. The variation of stress and temperature during the maturation and geological process results in the formation of micro- and macro-fractures in shale. These natural fractures make the geomechanics and fluid flow study of shale more challenging but also have an important factor for in hydraulic fracturing stimulation and production from this type of formation. Because of its multi-components, the lamination, or the arrangement of layer interface, of shale often has a significant effect on the mechanical, acoustic and anisotropic properties of shale (Al-Qahtani and Tutuncu, 2017).

---

[1] binhbt@pvu.edu.vn

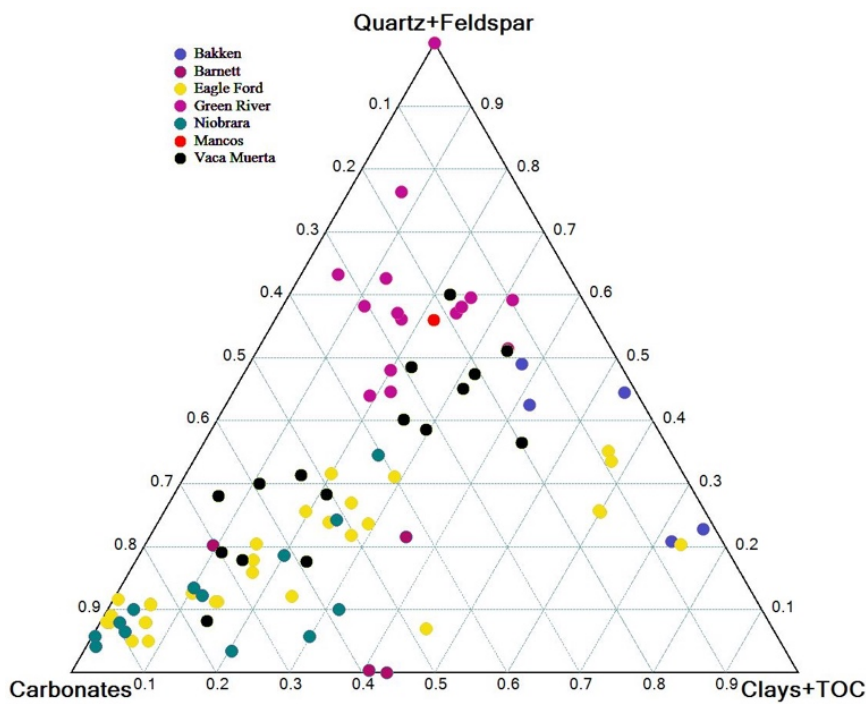

**Figure 1. Mineralogical composition of some common shale formations.**

The term "unconventional geomechanics" is used to describe the geomechanical study of unconventional formations. The main distinction between conventional geomechanics and unconventional geomechanics are in the inelastic shale matrix, stress sensitivity, low permeability, the fluid-rock interaction, and the presence of and natural fractures as well as the mechanical anisotropic characteristics of shale. The effect of the interaction between fluid and rock in conventional reservoirs is less significant compared that in unconventional reservoirs. The relatively higher pore size in conventional reservoirs allow the fluid to move in or out of the pore space quickly to reach the steady state when stress change. However, in unconventional reservoirs the movement of the fluid is the pore space is at much lower velocity resulting in the transient interaction. Also due to its very low permeability, the transport and storage properties as well as production of shale depend strongly on the stress (**Figure 2**) and mechanical interaction. In addition, high surface energy of the clay associated with shale complicates the interaction between fluid and rock enhancing the role of fluid, especially fluid electrochemical properties, on shale deformation. These enhance the role of geomechanics in any aspect of shale development project.

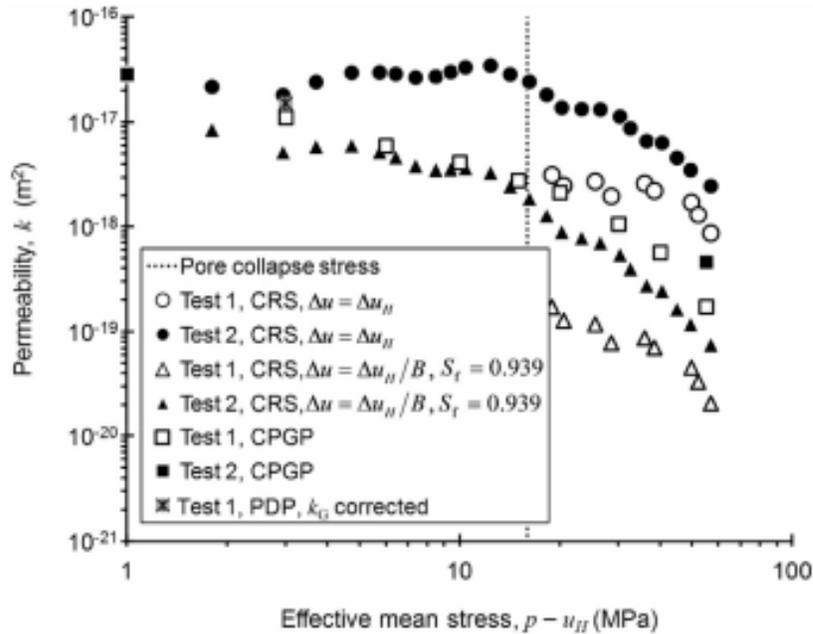

**Figure 2.** Stress dependence permeability and porosity of shale (Gutierrez 2014).

Shale mechanical properties can be measured in the laboratory and inside the wellbore using acoustic longing tools. While low frequency (static) properties obtained from tri-axial experiments may be relevant to field operation. and provide the most realistic mechanical data of shale, this method is rather expensive and limited to the number and location of collected samples. Hence, static data from triaxial experiments are used to calibrate the dynamic data obtained acoustic log. Acoustic logs measure the mechanical and acoustic properties of rocks at approximately 20 KHz or at ultrasonic frequency (>1MHz). Contrast to the low strain-rate experiments of static measurement in geomechanics laboratory, dynamic data from acoustic measurement depend on several factors that affect the propagation of energy. When acoustic waves propagate through a shale, the high frequency vibration of the transmitter creates the oscillatory motion of solid grain and the fluid in the pore space. Under rapidly oscillating deformations, the pore fluids do not have sufficient time to flow into low pressure regions, the rock will act as if it is unrelaxed or undrained. This means that the medium will behave stiffer in the unrelaxed state resulting velocity dispersion. On the other hand, if time is sufficient for fluid pressure to reach equilibrium, then the relaxed properties are measured as in the low frequency measurements. The behavior of shale under high frequency deformation depends on not only its fluid and rock properties such as mechanical properties, porosity, permeability, saturation, mineralogy, pore structures, density and viscosity but also on external parameters such as stress, temperature, and pore pressure. More importantly for shale, the electrochemical characteristics of the fluids inside the pore space, the fluid-shale interaction, the conductivity of shale and the presence of fractures have a considerable effect. An interpretation process is used to obtain static properties from dynamic log data that requires the knowledge of fluid and rock interaction. These makes the interpretation of acoustic data more challenging.

The most important mechanical properties of the formation that must be qualified for an unconventional resource development are Young's modulus, Poisson's ratio, shear modulus, compressive strength (typically unconfined compressive strength-UCS), tensile strength and the failure characteristics such as friction coefficient and cohesive strength. While Young's modulus measures the stiffness of shale under compression, shear modulus measured the resistance of shale to shear stress. Shale dynamic and static Young's modulus is typically less than 13 Mpsi (**Figure 3**) and shear modulus is often less than 5 Mpsi. The compressional wave velocity of shale varies from 4000 ft/sec to 10000 ft/sec while the shear wave velocity is typically less than 8500 ft/sec. The bulk modulus of shale, measured the resistance to volumetric compression, is typically less than 02. Mpsi. Poisson's ratio, determining the lateral expansion perpendicular to the direction of compression, of shale typically varies 0.2 to 0.4. Unconfined compressive strength is the maximum compressive stress that material can sustain before failure in uniaxial experiment. The unconfined compressive strength of shale is often less than 15000 psi. The maximum tension stress that shale can sustain before failure, or tensile strength, is typically less than 1500 psi, in most case this value is in the range of 300 to 800 psi. In hydraulic fracturing simulation and wellbore integrity analysis, shale failure models must be used. Shale can be failed under tension (tensile failure) or compression (shear failure). The failure of shale under tension occurs when tensile stress excesses its tensile strength. The failure of shale under compression is rather complex since under different confining stress, shale fails at different axial stress. Hence, different models have been proposed to predict the failure of shale. The Shale Young's modulus and Poisson's ratio are indispensable parameters in wellbore integrity analysis, hydraulic fracturing calculation, and reservoir modelling. For conventional reservoir, failure characteristics is used mostly in wellbore stability analysis to determine the optimal mud weight windows. However, these properties are used much widely for shale formations, especially in hydraulic fracturing simulation. In addition, Biot's coefficient, correlates the mean stress acting on shale grain (effective stress) with total stress and pore pressure, is also an important parameter used in all geomechanical calculation. Havens (2012) showed that the Biot's coefficient of the For Bakken formation varies from to 0.15 to 0.75. **Table 1** summarizes our laboratory experimental data from various formations.

**Table 1. Static Young's modulus and Poisson's ratio of common shale formations**

| Mechanical property | Eagle Ford | Bakken | Barnett | Niobrara | Vaca Muerta |
|---|---|---|---|---|---|
| Static Young's modulus (Mpsi) | 3-8 | 3.5-12 | 3-12 | 2-4 | 3.5-5.5 |
| Static Poisson's ratio | 0.15-0.35 | 0.3-0.4 | 0.2-0.3 | 0.11-0.26 | 0.1-0.25 |
| UCS (Mpsi) | 0.15-0.3 | 0.4-0.45 | 0.2-0.4 | 0.3-0.8 | 0.1-0.3 |
| Tensile strength (kpsi) | 0.3-0.8 | 1.0-2.0 | 0.3-0.5 | 0.4-1.5 | 0.5-1.0 |

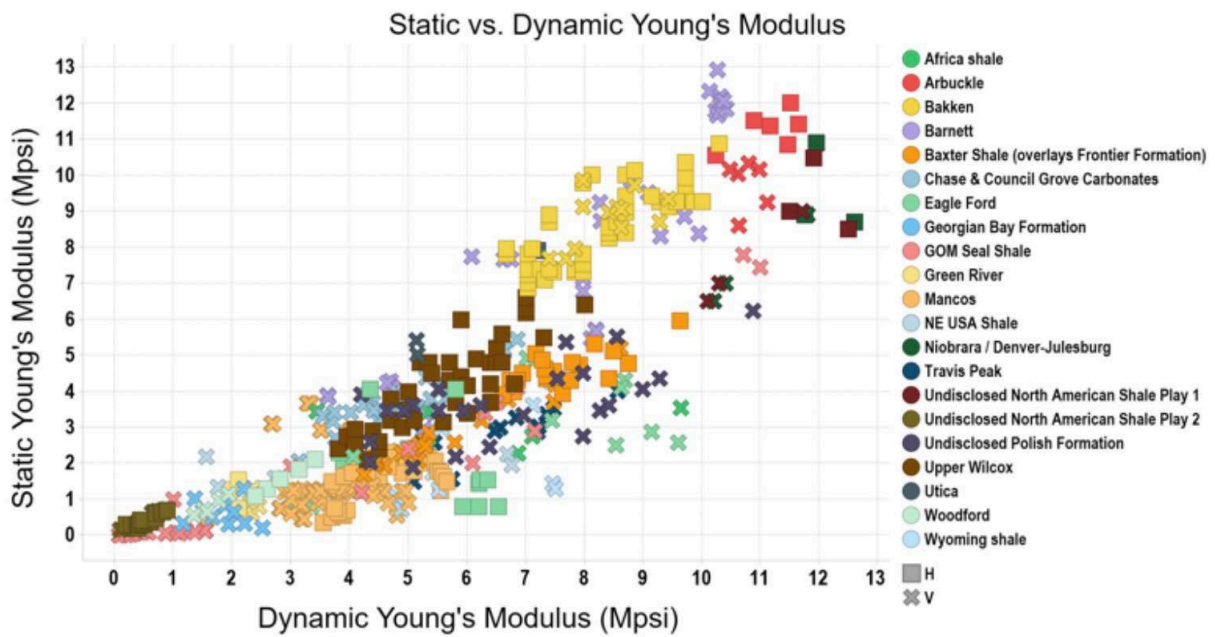

**Figure 3. Static and dynamic Young's modulus of common shale formations (Wehbe, 2022).**

The mechanical properties of shale are often anisotropic and considered to be transversely anisotropic media. It means that the properties of shale are realatively the same in one horizon but changes significantly in the direction normal to it. Due to its lamination, fabric structure and micro fractures, shale mechanical properties may significantly change with bedding direction. The distribution of clay and organic matter also determines the level of elastic anisotropy in shale reservoir formations. The kerogen maturity and bedding orientation are among the key parameters controlling shale's mechanical anisotropy. The local tectonic history also affects the elastic properties of shale formations. The relationship between elastic anisotropy and clay and kerogen content indicates that as clay and kerogen increase the amount of elastic anisotropy increases. The ratio of vertical to horizontal Young's modulus of shale can vary from formation to formation, for example this ratio for the Niobrara formation is typically from 1-1.5 (Bridges 2016). The mechanical anisotropy of shale affects many aspects of mechanical modelling and creates more challenges for numerical simulation as well as laboratory characterization. All related geomechanical calculations such as wellbore integrity analysis, hydraulic fracturing simulation, and reservoir modelling have to accounted for shale anisotropy. Anisotropy affects the deformation of shale and the induced pore pressure altering the effective stress state resulting in wellbore instability. The effect of mechanical anisotropy on wellbore stability increases with the increases of stress anisotropy of the formation (Aoki et al. 1993). Ghassemi (2016) suggests that fracture toughness promotes the growth of inner region fractures, and shale anisotropy affects the stress shadows, direction of hydraulic fractures, and the geometry of fractures. Shale mechanical anisotropy affects the performance of the reservoirs indirectly through its effect on hydraulic fracture propagation and geometry. The mechanical anisotropy may have important contribution to production performance due to its effect on permeability anisotropy evolution during production and proppant embedment. Although shale anisotropy is included in reservoir modeling, the effect of mechanical anisotropy on reservoir performance has not been comprehensively investigated in the reservoir engineering literature.

In addition to its in inelastic characteristics, the stress and fluid sensitivity of shale formations have been well recognized since the early days of conventional reservoir development as a key data for wellbore integrity analysis. Shale formations are also highly heterogeneous with the organic matter and compositional variations throughout the areal extent of the reservoirs. The level of maturity of the organic matter also influences the mechanical, acoustic, petrophysical and failure properties of organic rich shale formations. The mineralogical composition typically deviates from carbonate rich to quartz rich with clay and organic matter amount and distribution heterogeneity in the reservoir. Rock evaluation analysis is a type of bulk analyses that does not provide assessment of heterogeneity at small scales that are essential for better understanding of the coupled geomechanical and flow characteristics of the reservoir to achieve highest production potential.

**Geomechanics in reservoir characterization**

The most important geomechanical parameters for shale formation characterization are the in-situ stresses, namely vertical stress, maximum and minimum horizontal stresses. Although stress anisotropy is a misnomer because anisotropy implies the directional dependence of the material property, stresses are typically different in different directions, and stress is not a property, but a state. The term stress anisotropy refers to the difference between minimum and maximum horizontal stresses. The vertical stress can be obtained by integrating the density log. To determine the minimum horizontal stress a small injection test, such as mini-frac test, is typically conducted. In this test, small amount of fluid is injected into the formation until a small fracture is initiated. Then, the well is shut in to monitor fracture closure. During the test, pressure is recorded, and fracture closer pressure is considered to be the minimum horizontal stress. The maximum horizontal stress is typically obtained from the vertical and minimum horizontal stress as well as geomechanical properties. The direct determination of maximum horizontal fracture is rather more complex. Many methods have been proposed to determine this stress such as using wellbore breakout model (Zoback et al. 1985), stress polygon (Zoback 2010), micro-seismic focal mechanisms (Agharazi 2016), and borehole sonic measurement (Sinha et al. 2016). In-situ stress anisotropy is accounted in all unconventional geomechanics calculation. It attributes to the propagation and geometry of hydraulic fractures and has a long-term effect on reservoir performance (Bahrami et al., 2010).

Due to its very low permeability and porosity, not all parts of the formation are economically producible, especially at low oil price. This requires more advanced technology to be able to pinpoint the best production potential areas, or sweet spots. Sweet spots are identified by source-rock maturation and total organic carbon (TOC) content, formation thickness, natural fractures, and brittleness. The knowledge of geomechanics and geomechanical data are valuable information to identify sweet spots through core analysis and logging data, particular seismic data. The sweet spots are typically characterized by high TOC content, high gamma ray, high Young's modulus, low Poisson's ratio, low density, and low compressional velocity. Seismic data can be used to locate the reservoir location and depth and estimate the structure of the reservoir. Having the correlation between TOC and shale mechanical properties, seismic data can identify the presence, geometry, and TOC content source rock to identify the sweet spot. Although the sweet spot has some potential pitfall and even unrealizable as pointed out by Haskett (2014), more efforts should be spent to correlate the production potential of shale with its mechanical and maturation characteristics.

**Geomechanics in well construction**

Geomechanics knowledge and data are the contributing factors enabling the industry to drill longer horizontal wells in shorter amount of time. The drilling time has reduced considerably from few weeks or even months to less than two weeks today. This significant time reduction has the contribution of geomechanics, mainly accounted in wellbore stability and trajectory steering that require the knowledge of formation anisotropy, laminations, natural fractures and bedding planes. Wellbore stability is the main challenge when drilling in shale formations due to a number of technical issues such as the swelling of the shale, the interaction of drilling fluid and shale, and the bedding layer and microfracture characteristics of shale. Resolving these issues requires the geomechanical knowledge, especially fluid-shale interaction. With about two million wells in unconventional reservoirs, wellbore integrity analysis has become an important calculation in any unconventional reservoir development project. This requires a better understanding and modeling of the interactions between drilling or completion fluids and shale formation.

The most common instability of wellbore in shale formation is from its bedding and layering characteristics. Liu et al. (2016) and Dokhania et al. (2016) suggested the directions of bedding planes have a significant influence on wellbore instability. Other important factors that affect the integrity of wellbore are stress anisotropy and mechanical anisotropy. The mechanical anisotropy of shale is also found to be a very important factor in the stability of wellbore in shale formation (Dokhania et al. 2016; Li and Weijermars 2019). They found that when the in-situ stress anisotropy increases, the breakdown and collapse pressure decrease, and the safe drilling window decreases gradually. The increasing of mechanical anisotropy decreases the breakdown and collapse pressure narrowing of the safe mud weight windows. The other challenging drilling in shale formation is the present of natural fractures and the bedding that result in the imbibition of drilling fluid deeper along shale layering interfaces changing the mechanical properties of shales potentially causing swelling or disintegration of shale matrix and resulting in time dependent well integrity.

Because shale very small pore size increases its membrane efficiency, osmosis process becomes more important in wellbore stability analysis. For conventional reservoirs, the pore size is significantly larger than the diameter of the solute molecules, and the membrane coefficient is very small. Hence, the effect of osmosis pressure is negligible. However, shale pore throat size is not too large compared to the solute molecule size resulting in higher membrane efficiency, typically less than 10%. This small pore size prevents the transport of the solute in and out of the shale resulting the increase of osmosis pressure changing the pore pressure and causing the failure of shale facilities the formation of secondary micro-fractures inside shale. The failure of the shale and the formation of the micro-fracture due to the invasion of fluids is called fluid-induced wellbore instability. Therefore, the common wellbore stability analysis approach ignoring the shale-fluid interaction may underestimate the required mud weight to prevent wellbore collapse (Dokhania et al. 2016). For wellbore stability analysis in shale, coupled fluid flow and geomechanics models are often employed to estimate the optimal mud weight as well as drilling fluid salinity to deal with this fluid-induced wellbore instability problem. For hydrocarbon bearing shale formations, with the presence of hydrocarbon phase, the transport and surface properties of shale such as matrix permeability and wettability become the governing factors for the transport of drilling fluid into the matrix affecting the stability of the wellbore. Hence, for hydrocarbon bearing shale formations, multiphase transport models should be used to replace the single-phase transport models (Bui and Tutuncu 2018).

**Geomechanics in well stimulation**

The most important contribution of geomechanics on the development of unconventional resources is in hydraulic fracturing operations and modelling. Horizontal well and hydraulic fracturing are the main technologies that create the shale revolution. While oil well fracturing technology has been available for many years since the first successful fracturing jobs in 1949 at Stephens County, Oklahoma, and Archer County, Texas by Halliburton, horizontal drilling is the main technology advancement enabling the success of fracturing in unconventional reservoirs. The success of hydraulic fracturing operations depends heavily on the knowledge of the rock properties as well as the in-situ stress regime. Without the knowledge of the in-situ stress, the horizontal wellbore may be drilled in the direction parallel to maximum horizontal stress and resulted in transverse hydraulic fracture that propagates along the wellbore. This fracture pattern creates less fractured area as well as stimulated volume than the fractures propagates perpendicular to the wellbore. Hence, the direction of minimum horizontal stress is a critical data to design the direction of horizontal well to maximize the facture surface and increase the recovery efficiency.

The determination of perforation depth, location of fracture, number of stages is also obtained from geomechanics data and calculation. The in-situ stress and geomechanical properties of formation are used to design the perforation size and location as well as number of perforations per cluster, space between perforation clusters, perforation density to success fully initiate the hydraulic fractures. Without the knowledge of stress field and geomechanical properties formation, the design of clusters and perforation may not be proper resulting in the inequal flow of the fracturing fluid into clusters. This means some clusters are not stimulated during hydraulic fracturing treatment creating non stimulated zone in the micro-seismic observation (**Figure 4**) and non-producing clusters (**Figure 5**). The analysis of production logs from more than 100 horizontal wells by Millers et al. (2011) suggested that two-thirds of gas production is from only one-third of clusters in some formation suggesting that some clusters are not sufficiently stimulated. Since fluid preferably flow toward lower stress interval, the number of perforations is reduced in lower-stress interval and increase in higher-stress interval to balance the flow rate for each cluster (Wutherich and Walker, 2012). This ensures the equal stimulation of each interval and avoid non-producing clusters improving hydrocarbon recovery.

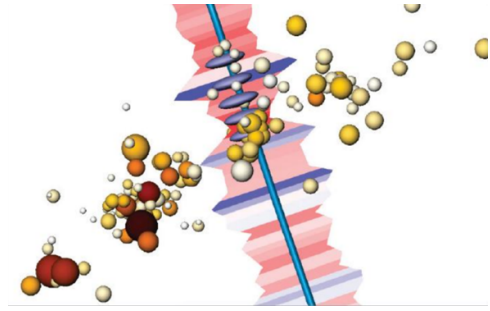

**Figure 4.** Micro-seismic event concentrated more near the lower-stress interval (red color) (Wutherich and Walker 2012).

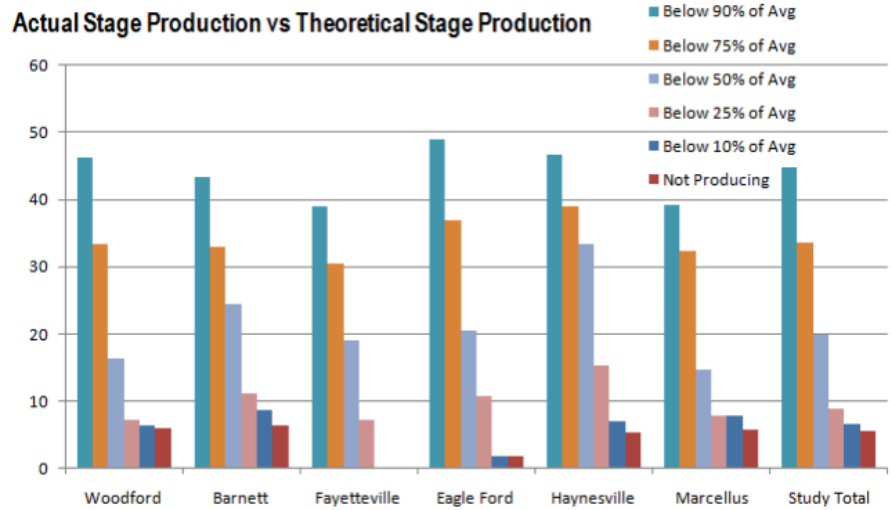

**Figure 5.** Actual stage production versus theoretical stage production (Miller et al. 2011).

In the early days of hydraulic fracturing, simple planar fracture geometry is often assumed. In highly anisotropic horizontal stress field, hydraulic fractures are typically planar extending far away from the wellbore. While more complex fractures networks are typically observed in lower horizontal stress anisotropy formations. The fracture geometry also depends on the presence of the natural fractures and their interaction with hydraulic fracturing. Today, with the advanced of geomechanics more complex model capturing the complexity of the fracture network. This enable more accurate simulation of the reservoir improving hydrocarbon recovery. Along with fracturing simulation, the determination of stimulated reservoir volume (SRV) from seismic data also provide more accurate evaluation of the success of fracturing operation and provide a validation for mechanical earth model.

Young's modulus and Poisson's ratio are used to determine the geometry and dimensions of hydraulic fractures. These parameters are also used to classify shale as brittle and ductile. Ductile shales typically have high Poisson's ratio and low Young's modulus while brittle shales typically have low Poisson's ratio and high Young's modulus. The complexity of hydraulic fracture network increases from ductile shales to brittle shales. Secondary fractures often are often observed for brittle shales. In unconventional geomechanics literature, the brittleness index is a widely used concept although there are different definitions and well as several models to determine this index. Based on brittleness index, shales are classified as. Brittleness index is a practical parameter used in determination of the hydraulic fracture geometry (**Figure 6**) and hydraulic fracturing fluid selection (**Figure 7**).

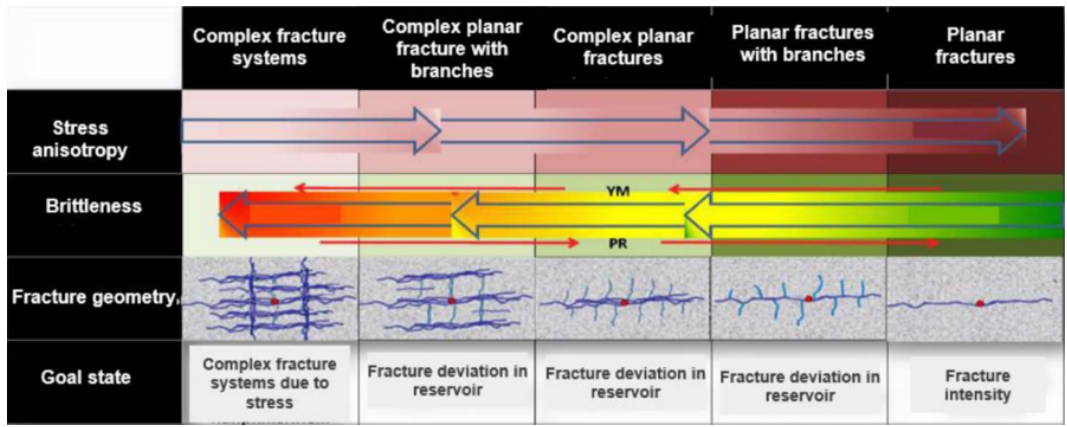

Figure 6. Effect of shale brittleness on hydraulic fracture geometry (Nenasheva et al. 2018).

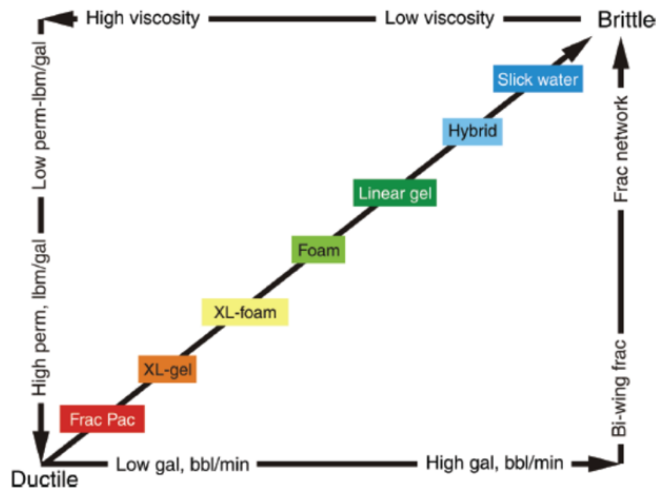

Figure 7. Effect of shale brittleness on fracturing fluid selection (Chong et al. 2013).

In hydraulic fracturing operations, identify the fracturing zone and determination of optimize wellbore and fracture spacing still remains challenging. The in-situ stresses are also used in determination of optimal stage length and well spacing. In fracture optimization, identifying fracture interval, fracture designing, and predicting facture geometries are all based on geomechanical properties of shale. In-situ stress magnitude and direction are used in determining the direction of wellbore to avoid wellbore instability problems and to maximize stimulated reservoir volume. Horizontal wellbores are recommended to drilled in the direction of the maximum horizontal stress. This helps to create the vertical hydraulic fractures perpendicular to the wellbore and results in higher fracture surface area as well as larger stimulated reservoir volume. Also, the breakdown pressure, pressure to initiate fracture, is also lower for the wellbore drilled in the minimum horizontal stress direction. Placing hydraulic factures too far away from each other may result in poorly stimulated reservoir. However, fracture spacing to small can result in stress shadowing effect. Stress shadowing is the variation of the local in situ stress induced by the previous hydraulic fractures than affect the propagation of the hydraulic new fractures. Due to the stress shadowing effect, the minimum horizontal stress tends to increase locally that contains the formation of pressure in the direction perpendicular to the direction of horizontal stress and hydraulic fractures tend to propagate vertically out of the target zone depending on the magnitude of the vertical stress. Dohmen et al. (2014) suggested that stress shadowing modeling and micro-seismic monitoring can be used as tools to optimize the fracture spacing.

**Geomechanics in reservoir modeling and management**

With the development of new tools for fracture characterization using seismic data, seismic-driven reservoir simulation and monitoring are becoming the standard for improving reservoir description (Ouenes et al. 2004; Li et al. 2014; Ramanathan et al. 2014). The understanding of geomechanics and seismology have provided a tool for modeling the fracture geometry and integration of the seismic and stimulation data in the reservoir modeling and production forecasting. In reservoir modeling development of the geologic model with representative formation parameters is critical for the accurate of simulation and production forecasting. The estimation of the reservoir properties such as fracture conductivity, porosity, permeability is the most important contribution of geomechanics on unconventional reservoir modeling. The transport properties of the formation are estimated from the mechanical failure of rock to obtain the formation of discrete fracture network (DFN). The DFN model is then validated with the micro-seismic data. The determination of stimulated reservoir volume is also one the important contribution of geomechanics on reservoir modeling and hydraulic fracturing evaluation. The SRV concept was introduced by Fisher et al. (2004) to relate production performance using micro-seismic data collected during Barnett shale hydraulic fracturing operations. The estimation of SRV from micro-seismic mapping can be correlated to well performance, drainage volume, and ultimate recovery. Mayerhofer et al. (2010) suggested that the size of the SRV depends\ on natural fractures, fracture spacing, formation thickness, in-situ stress, rock mechanical properties (brittleness) and the geological characteristics of the formation. Although SRV may not realistically represent the real production enhancement volume as Cipolla and Wallace (2014) suggested, it provide a preliminary estimation of the fracture distribution and conductivity and can be used as an easement of the hydraulic fracturing performance.

The presence of natural fractures, micro- and macro-fractures, making the study of geomechanics for unconventional reservoirs more challenging. The hydraulic fracture propagation path is highly affected by the interactions between the natural fractures and hydraulic fractures, fracturing fluid, and proppant and fluid selection. Local anisotropic rock properties and reservoir mineralogy and natural fractures are the key parameters determining fluid transport and proppant placement, associated in situ stress alterations during fracturing process, and production in unconventional tight oil and shale gas reservoirs. Excluding these important factors in this naturally fractured tight organic-rich formations results in planar symmetric bi-wing fractures as in high permeability conventional reservoirs. This make the 3D reservoir model unreliable for reservoir simulation resulting in misleading production forecasting and reservoir management plans. However, the recent understanding of geomechanics especially the knowledge of the interaction between natural and hydraulic fractures has helped to overcome this limitation. Today DFN model become widely used as a standard in the unconventional reservoir modeling to represent the complexity of the hydraulically fractured reservoir. DFN model comprises the fracture network complexity providing a detailed representation of the fracture network. DFN model can be developed using a combination of deterministic, directly imaged seismic, imaging logs; full wave dipole sonic logs; local and regional geological data; and seismic surveys. This marks a significant contribution of the geomechanics in unconventional reservoir modeling and management.

Coupling fluid flow model to geomechanics model for modeling of fluid and rock interaction has received much attention of the petroleum industry. Petroleum engineers often deal with many problems that involve complex interactions between geomechanics, fluid flow, and heat transfer. Hence, they often have to solve the geomechanical model together with the fluid flow model to evaluate the effect of rock deformation/failure on the hydrocarbon production. One of the primary objectives of the coupled fluid flow and geomechanics modeling is to account for the effect of rock deformation on the flow and associated mechanical interaction of the reservoir. The fluid flow equation is related to geomechanics equation by the volumetric strain representing the volumetric variation of rock due to pressure, stress, and temperature. The flow properties of the reservoir, particularly permeability, is often related to the volumetric variation of rock due to fluid pressure, in-situ stress, and temperature. Because of the present of the hydraulic fractures and these fractures apertures change with the variation of stress after hydraulic fracturing and during production. Phenomena such as proppant embedment continuously alter the facture transport properties affecting the performance of the well. Hence, the prediction of unconventional reservoir performance is impossible without coupled fluid flow and geomechanics modeling. Coupled models have been developed and used intensively in not only hydraulic fracturing simulation but also in reservoir modeling and production forecasting. Kim and Moridis (2012) presented a coupled geomechanics and fluid flow model using multiple porosity model for shale reservoirs. Their results suggested that coupled fluid flow and geomechanics models using the double or multiple porosity model is more appropriate for modeling of shale gas reservoir than using the uncoupled flow model. Fakcharoenphol et al. (2013) suggested that water-induced stress is one of the mechanisms for enhancing formation permeability and hence improving gas recovery. The coupled modeling approaches have shown some potential in solving not only complex reservoir modeling problems but also in hydraulic fracturing simulation as well as wellbore integrity analysis. However, the challenge is still coming from the complexity of the model as well as complex of the physics in the reservoir that should be accounted for. Hence one of the drawbacks of coupled simulation is the computational time and cost that limit the scale of investigation for large scale reservoirs.

**Geomechanics in sustainability development**

One of the environmental problems associated with unconventional reservoir development is the failure of the cement sheath that results in the mitigation of the hydrocarbon to the upper water aquifers and to the surface. The sustainability of the cement sheath during the lifecycle of wellbore, especially during hydraulic fracturing, is also an important aspect for sustainability development that require a comprehensive understanding of geomechanics. The leaking of the hydrocarbon along the wellbore in the space between casing and rock due to the weak integrity of cement sheath is a serious concern as the main reason for contaminating the underground water and damaging the environment. The failure of the cement sheath is not only associated with the shrinkage and contamination of the cement but also with the high internal pressure used in hydraulic fracturing operation and the change of temperature. The most common failure of the cement in unconventional reservoirs is the result of the excessive internal pressure of inside the wellbore that creates the microcrack and the loss integrity in cement sheath. Geomechanics studies reveal that the integrity of this cement strongly depends on mechanical properties of cement wellbore geometry, rock mechanical properties (Thiercelin et al. 1998), especially the tensile strength of cement and in-situ stress (Bui and Tutuncu 2014), the change of temperature (Dusseault et al. 2000), and the eccentricity of casing in wellbore (Lui et al. 2018). Hence, it is commonly recommended to increase the tensile strength of cement the thickness of the casing to improve it integrity.

Geomechanics is also in important tool in unconventional reservoir sustainability development for prediction of seismic emission and earthquake associated with fluid injection. Large gas shale and tight oil reserves with significant unconventional development activities have been geographically located in the areas with minor seismicity and considered to have small potential for possible earthquakes. In the last decade there has been significant growth in hydraulic fracturing operations and associated micro-seismic monitoring as a result of increased number of operations in shale reservoirs and the associated production. Recent sizable earthquakes (2–5.3 Richter scale magnitudes) in the states such as Texas, Oklahoma, Colorado, Pennsylvania, and Ohio have raised further concerns and associated interest in the role the hydraulic fracturing injection and hydrocarbon production on induced seismicity (Rutqvist et al., 2013, McGarr, 2014, Tutuncu and Bui 2016). During fluid injection operations, if the injection site is close to a fault, fluid is forced along the fault discontinuity relieving the stress acting on the fault altering stress state and as a result reactivation of the fault may be imminent that most often triggers seismic-induced events. In general, micro scale earthquakes concentrate in the reservoir intervals and where pressure gradients are the largest, at the sealing faults and other barriers, or along the propagating fractures within the reservoir. The micro-seismic events are typically considered to be shear failures that occur around the opening of a tensile hydraulic fracture as the fracture propagates. Depending on the extent of the fault slide and/or rupture size of the formation, the level of seismicity typically increases.

The occurrence of seismic events is associated with the deformation and failure of rock due to the change of pore fluid pressure. Hence, to predict or determine the source of these events, coupled fluid flow and geomechanics provide the best solution. Tutuncu and Bui (2016) conducted a numerical simulation to determine the effect of injected fluid on the stress change along the fault in the reservoir. Pore pressure alteration and associated changes in the in-situ stress state were calculated using a coupled geomechanics and fluid flow model to predict the potential for induced seismicity occurrence in the field. The induced seismicity was calculated using the complex shear slippage on the fault plane. They suggested that the distance to the fault, fault geometry (including fault orientation) and fault geomechanical properties are among the critical parameters determining released energy and associated induced seismicity occurrence. Fluid injection rate, injection fluid viscosity and formation permeability are also among the key parameters for induced seismicity assessment. In addition to the local geology plays and fault characteristics, high injection rates create high potential for induced seismicity events. They also emphasize on the understanding of the interactions between the injected fluid and the fault plane. These results suggested that coupled modeling can be a very useful tool for assessment of seismic events associated with unconventional reservoir operations.

**Fluid and rock interaction**

Understanding the interaction between fluid and rock is one important step toward shale reservoir characterization to determine the reliable formation properties. Formation properties are significantly changed when different fluids are introduced. During the hydraulic fracturing stimulation, a large volume of water and chemical are injected into the formation, typically from 1000 to 5000 SCF/ft. They interaction with the formation and change the its mechanical properties. The interpretation of seismic data for unconventional reservoir is rather more complex than that for conventional reservoir because of the new fluids introduced to the formation during hydraulic fracturing. During hydraulic fracturing a larger volume of water is injected to formation. The injected fluids interact with not only formation fluids but also with the sale matrix alternating the mechanical behavior of rock affecting it deformation and failure as well as hydrocarbon recovery. Due to the small grain size and the strong surface electrochemical properties of shale grains, the effect of fluid on mechanical properties and deformation of shales is more significant than for unconventional reservoir. The static mechanical properties of shale were reported to change significantly change with water saturation and water chemical properties (Zhang et al. 2006), especially the dynamic properties (Adekunle et al. 2022). The experiment by Lai et al. (2016) suggested that the reduction of shear velocity is more significant than that of compressional velocity.

The swelling of shale when contact with injected water is also an important aspect that geomechanics can contribute to the development of unconventional reservoir. With the oil recovery factor ranging from 4 to 7%, porosity typically from 5 to 10%, and initial oil saturation typically less than 80%, the volume of oil recovered from unconventional reservoir is typically less than 0.5% of the reservoir bulk volume. This volumetric depletion can be less than the volumetric expansion of shale due to swelling, especially for shale with high smectite content. The swelling of shale forcing oil out of shale matrix and fractures but also reduces the permeability of matrix and fractures, hence, causes a significant reduction of the oil production from the unconventional reservoir (Bui and Tutuncu, 2018).

Due to very small pore size of shale, typically range from 5 to 20 nm, the gas inside the pore throat condenses into liquid form. This phenomenon is known as capillary condensation have several effects production as well as transport and mechanical characteristics of shale. The high capillary pressure in shale significantly shifts thermodynamic properties, including phase compositions and the dew-point pressure (Shapiro et al., 2000). The shift of the phase envelope due to nanopores affects the evaluation of the hydrocarbon in place (Ambrose et al., 2012, Didar and Akkutlu, 2013) and production decline of shale gas formations (Nojabaei et al., 2013). The condensation of the gas inside nanopores may also affect the acoustic properties of shale formations. Bui and Tutuncu (2015) suggested that the change of acoustic data for the same formation at different time during the lifecycle of the reservoir is the result of the change of the phase behavior of the fluid in during the production. Capillary condensation has a stronger effect on acoustic properties of shale in high frequency range.

**Summary and Remarks**

Conventional and unconventional geomechanics approaches have greatly contribute to the success of the unconventional resource development. Geomechanical data are indispensable for unconventional reservoir projects and have been used intensively in all aspects from formation characterization and well construction to hydraulic fracturing and reservoir modelling as well environmental assessment and sustainability development. In reservoir characterization, mechanical anisotropy, natural fractures, and in-situ stresses are extensively investigated in the literature with the recent focus on sweet spot identification. In addition to natural fractures, bedding characteristic, the inelastic and anisotropic mechanical properties of shale, the coupled transport and fluid and rock interaction has received more attention in well construction. The focus of geomechanics in well stimulation, particularly hydraulic fracturing, is on optimizing the wellbore and hydraulic fracture spacing as well as long term behavior of proppant and fracture interaction. In reservoir modelling and management, multiple coupled modeling and integration of seismic data into reservoir simulator are an active area of research with the focus on simplified approach to reduces the computational cost and time. The physical phenomena that has profound effect on unconventional reservoir production such as capillary condensation and fluid-rock interaction are also important to understand how the mechanical aspect at nano-scale can affect reservoir long-term performance. In addition, predicting micro-seismic events such as earthquake is an important contribution geomechanics to environmental protection and sustainability development.

**Reference**

1. Adekunle O., Bui B., and Katsuki D. Effects of Chemical Osmosis on Swelling Clays and the Elastic Properties of the Pierre Shale with Its Implications for Oil Recovery. *International Journal of Rock Mechanics and Mining Sciences* **155:** 105110.
2. Agharaz A. Determining Maximum Horizontal Stress with Microseismic Focal Mechanisms-Case Studies in the Marcellus, Eagle Ford, Wolfcamp. Paper URTeC-2461621 presented at the Unconventional Resources Technology Conference held in San Antonio, Texas, USA.
3. Al-Qahtani A.A. and Tutuncu A.N. 2017. Qualitative and Quantitative Impact of Thin Rock Lamination on Acoustic and Geomechanical Properties of Organic-Rich Shale. Paper SPE-183806-MS presented at the SPE Middle East Oil & Gas Show and Conference, Manama, Bahrain.
4. Aoki T., Tan C.P., and Bamford W.E., 1993. Effects of Deformation and Strength Anisotropy on Borehole Failures in Saturated Shales. *International Journal of Rock Mechanics and Mining Sciences & Geomechanics Abstracts*. **30** (7): 1031–1034.

5. Bahrami H., Rezaee R., and Asadi M.S. 2010. Stress Anisotropy, Long-Term Reservoir Flow Regimes and Production Performance in Tight Gas Reservoirs. Paper SPE-136532-MS presented at the SPE Eastern Regional Meeting, Morgantown, West Virginia, USA.
6. Bridges M. 2016. Mechanical properties of the Niobrara. M.Sc. thesis, Colorado School of Mines, Golden, Colorado, USA.
7. Bui B.T. and Tutuncu A. N., 2018, Modeling the Swelling of Shale Matrix in Unconventional Reservoirs. *Journal of Petroleum Science and Engineering*, **165**: 596–615.
8. Bui B.T. and Tutuncu A.N. 2018. A Coupled Geomechanical and Flow Model for Evaluating the Impact of Drilling Fluid Imbibition in Reservoir Shale Formations. Paper ARMA-2018-075 presented at the 52$^{nd}$ U.S. Rock Mechanics/ Geomechanics Symposium, Seattle, Washington, USA.
9. Bui B.T. and Tutuncu A. N., 2015, Effect of Capillary Condensation on Geomechanical and Acoustic Properties of Shale Formations. *Journal of Natural Gas Science and Engineering*, **26**: 1213-1221.
10. Chong K.K., Grieser V.W., Passman A., Tamayo C.H., Modeland N., and Burke E.B. 2010. A Completions Guide Book to Shale-Play Development: A Review of Successful Approaches toward Shale-Play Stimulation in the Last Two Decades. Paper SPE-133874-MS presented at the Canadian Unconventional Resources and International Petroleum Conference, Calgary, Alberta, Canada.
11. Cipolla C. and Wallace J., 2014. Stimulated reservoir volume: A Misapplied Concept?. Paper SPE-168596-MS presented at the SPE Hydraulic Fracturing Technology Conference, Woodlands, Texas, USA.
12. Dohmen T., Zhang J., and Blangy JP. 2014. Measurement and Analysis of 3D Stress Shadowing Related to the Spacing of Hydraulic Fracturing in Unconventional Reservoirs. Paper SPE-170924-MS presented at the SPE Annual Technical Conference and Exhibition, Amsterdam, The Netherlands.
13. Dokhania V., Yu M., and Bloys B. 2016. A Wellbore Stability Model for Shale Formations: Accounting for Strength Anisotropy and Fluid Induced Instability. *Journal of Natural Gas Science and Engineering*, **32**: 174-184.
14. Dusseault M.B., Gray M.N., and Nawrocki P.A. 2002. Why Oilwells Leak: Cement Behavior and Long-Term Consequences. Paper SPE-64733-MS presented at International Oil and Gas Conference and Exhibition in China, Beijing, China.
15. EIA. 2013. Technically Recoverable Shale Oil and Shale Gas Resources: An assessment of 137 shale formations in 41 countries outside the United States. Technical report, U.S. Energy Information Administration.
16. EIA. 2014. Annual Energy Outlook 2014 with Projections to 2040. Technical report, U.S. Energy Information Administration.
17. EIA. 2018. Annual Energy Outlook 2019 with Projections to 2050. Technical report, U.S. Energy Information Administration.
18. Fakcharoenphol, P., Charoenwongsa, S., Kazemi, H., and Wu, S. 2013. The Effect of Water Induced Stress to Enhanced Hydrocarbon Recovery in Shale Reservoir. Paper SPE-158053-PA, *SPE Journal*, **3**(03): 897–909.
19. Fisher M.K., Heinze J.R., Harris C.D., Davidson B.M., Wright C.A., and Dunn K.P., 2004. Optimizing Horizontal Completion Techniques in the Barnett Shale Using Microseismic Fracture Mapping. Paper SPE-90051-MS presented at the SPE Annual Technical Conference and Exhibition, Houston, Texas, USA.
20. Ghassemi A. 2016. Impact of Fracture Interactions, Rock Anisotropy and Heterogeneity on Hydraulic Fracturing: Some Insights from Numerical Simulations. Paper ARMA-2016-283 presented at 50$^{th}$ U.S. Rock Mechanics/Geomechanics Symposium, Houston, Texas, USA.
21. Gutierrez M., Katsuki D., and Tutuncu A. N., 2014, Determination of the Continuous Stress-dependent Permeability, Compressibility and Poroelasticity of Shale, *Marine and Petroleum Geology*, **68**: 614-628.

22. Haskett J.W. 2014. The Myth of Sweet Spot Exploration. Paper SPE-170960-MS presented at the SPE Annual Technical Conference and Exhibition, Amsterdam, the Netherlands.
23. Havens J. 2012. Mechanical Properties of the Bakken Formation. M.Sc. thesis, Colorado School of Mines, Golden, Colorado, USA.
24. Kim, J. and Moridis, G. J. 2012. Numerical Studies on Coupled Flow and Geomechanics with the Multiple Porosity Model for Naturally Fractured Tight and Shale Gas Reservoirs. Paper ARMA-2012-296 presented at the 46th U.S. Rock Mechanics/Geomechanics Symposium, Chicago, Illinois, USA.
25. Lai B., Li H., Zhang J., and Jacobi D., and Georgi D. 2016. Water-Content Effects on Dynamic Elastic Properties of Organic-Rich Shale, Paper SPE-175040-PA, *SPE Journal,* **21** (02): 635 – 647.
26. Li Q., Zhmodik A., and Boskovic D. 2014. Geomechanical Characterization of an Unconventional Reservoir with Microseismic Fracture Monitoring Data and Unconventional Fracture Modeling. Paper SPE-171590-MS presented at the SPE/CSUR Unconventional Resources Conference, Calgary, Alberta, Canada.
27. Li W. and Weijermars R., 2019, Wellbore Stability Analysis in Transverse Isotropic Shales with Anisotropic Failure Criteria. *Journal of Petroleum Science and Engineering*, **176**: 982-993
28. Liu X., Zeng W., Liang X., and Lei W. 2016. Wellbore Stability Analysis for Horizontal Wells in Shale Formations. *Journal of Natural Gas Science and Engineering*, **31**:1-8.
29. Mayerhofer M.J., Lolon E., Warpinski N.R., Cipolla C.L., Walser D.W., and Rightmire C.M. 2010. What Is Stimulated Reservoir Volume? Paper SPE-119890-PA. *SPE Production & Operations*, **25** (01): 89-98.
30. McGarr A. 2014. Maximum Magnitude Earthquakes Potentially Induced by Fluid Injection. *Journal Geophysical Research: Solid Earth*, **119**: 1008-1019.
31. Miller K.C., Waters A.G., and Rylander I.E. 2011. Evaluation of Production Log Data from Horizontal Wells Drilled in Organic Shales paper SPE-144326-MS presented at the North American Unconventional Gas Conference and Exhibition, The Woodlands, Texas, USA.
32. Nenasheva M., Okunev M., Sleta N., Timirgalin A., Zhukov V., Garenskikh D., Volkov G., and Priklonsky O. 2018. The Best Practices and Approaches for Replication of Achimov Formation Development Technologies. Paper SPE-191473-18RPTC-MS presented at the SPE Russian Petroleum Technology Conference, Moscow, Russia.
33. Ouenes A., Zellou A., Robinson G., Balogh D., and Araktingi U. 2004. Seismically driven improved fractured reservoir characterization. Paper SPE-92031-MS presented at the SPE International Petroleum Conference in Mexico held in Puebla, Mexico.
34. Ramanathan V., Boskovic D., Zhmodik A., Li Q., and Ansarizadeh M. 2014. Back to the Future: Shale 2.0 - Returning back to Engineering and Modelling Hydraulic Fractures in Unconventionals with New Seismic to Stimulation Workflows. Paper SPE-171662-MS presented at the SPE/CSUR Unconventional Resources Conference, Calgary, Alberta, Canada.
35. Rutqvist J., Rinaldi A.P., Cappa F., and Moridis G.J. 2013 Modeling of Fault Reactivation and Induced Seismicity during Hydraulic Fracturing of Shale-gas Reservoirs. *Journal of Petroleum Science and Technology*, **107**: 31-44.
36. Sinha K.B., Walsh J.J., and Waters A.G. 2016. Determining Minimum and Maximum Horizontal Stress Magnitudes from Borehole Sonic Measurements in Organic Shales ARMA-2016-298 presented at the 50[th] U.S. Rock Mechanics/Geomechanics Symposium, Houston, Texas, USA.
37. Thiercelin M., Dargaud B., Baret J.F., and Rodriguez W.J. 1998. Cement Design Based on Cement Mechanical Response. Paper SPE-52890-PA, *SPE Drilling and Completion,* **13** (4): 266-273.

38. Tutuncu A.N. and Bui B.T. 2016. A Coupled Geomechanics and Fluid Flow Model for Induced Seismicity Prediction in Oil and Gas Operations and Geothermal Applications. *Journal of Natural Gas Science and Engineering*, **29**: 110-124.
39. Wehbe N. 2022. Anisotropic Dynamic and Static Geomechanical Property Correlations in Shale Formations. M.Sc. thesis, Colorado School of Mines, Golden, Colorado, USA.
40. Wutherich K. and Walker J.K. 2012. Designing Completions in Horizontal Shale Gas Wells: Perforation Strategies paper SPE-155485-MS presented at the SPE Americas Unconventional Resources Conference, Pittsburgh, Pennsylvania, USA.
41. Zhang J., Al-Bazali T., Chenevert M., Sharma M., and Clark D. 2006. Compressive Strength and Acoustic Properties Changes in Shale with Exposure to Water-Based Fluids. Paper ARMA-06-900 presented at the 41$^{st}$ U.S. Symposium on Rock Mechanics (USRMS), Golden, Colorado, USA.
42. Zoback M.D., Moss D., Mastin L., and Anderson R., 1985. Well Bore Breakouts and In Situ Stress. *Journal of Geophysical Research*, **90** (B7): 5523-5530.
43. Zoback M.D. 2010. Reservoir Geomechanics. Cambridge University Press, New York, USA.